\documentclass[showpacs,amsmath,amssymb]{revtex4}

\usepackage{graphicx}
\usepackage{dcolumn}
\usepackage{bm}
\usepackage{amsmath,epsfig}

\newcommand{\abs}[1]{\left\vert#1\right\vert}
\newcommand{\ket}[1]{\left\vert#1\right\rangle}
\newcommand{\bra}[1]{\left\langle#1\right\vert}
\newcommand{\braket}[2]{\left\langle#1\vert#2\right\rangle}
\newcommand{\etal}{\textsl{et. al.}}
\newcommand{\Name}[1]{#1,}
\newcommand{\REVIEW}[4]{\textsl{#1}, \textbf{#2} ({#3}) {#4}}
\newcommand{\Book}[1]{\textsl{#1}}
\newcommand{\Vol}[1]{Vol. \textbf{#1}}
\newcommand{\Publ}[1]{(#1)}
\newcommand{\Year}[1]{(#1)}

\begin{document}

\title{The non dissipative damping of the Rabi oscillations as a ``which-path'' information}

\author{M. Tumminello, A. Vaglica, and, and G. Vetri}

\affiliation{
  Istituto Nazionale di Fisica della Materia and Dipartimento di Scienze Fisiche ed Astronomiche dell'Universit\`{a} di Palermo-
  via Archirafi 36, 90123 Palermo, Italy} 

\date{\today}

\begin{abstract}
Rabi oscillations may be viewed as an interference phenomenon due
to a coherent superposition of different quantum paths, like in
the Young's two-slit experiment. The inclusion of the atomic
external variables causes a non dissipative damping of the Rabi
oscillations. More generally, the atomic translational dynamics
induces damping in the correlation functions which describe non
classical behaviors of the field and internal atomic variables,
leading to the separability of these two subsystems. We discuss on
the possibility of interpreting this intrinsic decoherence as a
``which-way" information effect and we apply to this case a
quantitative analysis of the complementarity relation as
introduced by Englert [Phys. Rev. Lett. \textbf{77}, 2154 (1996)].
\end{abstract}

\maketitle

Young's double-slit experiment may be considered as an emblematic
example to introduce the wave-particle duality, which, to quote
Feynman \cite{Fey}, is the basic mystery of Quantum Mechanics
(QM). This experiment is frequently discussed in the introductory
part of QM textbooks \cite{Coh} to explain how the interference
pattern and the which-way information on the particle trajectory
are mutually exclusive behaviors in QM. Moreover, this
wave-particle dualism belongs to the enlarged contest of
complementarity \cite{Scu, Zei}, which has its root in the
superposition principle. Coherent superposition of two states or,
more generally, of two quantum paths leads to interference effects
which may manifest itself as probability oscillations of some
``populations". For example, when a two-level atom interacts with
the field of an optical cavity, the atomic level populations
undergo to the well known Rabi oscillations. But, what is the
interference effect underlying this phenomenon? In other words,
which paths do interfere? \\ When the atomic external degrees of
freedom are included in the model, the internal dynamics of the
atom correlates with the translational variables. This information
transfer towards the atomic external variables is at the origin of
an intrinsic decoherence effect, that is, it causes a non
dissipative damping of the Rabi oscillations \cite{Vag1, Cus}. Non
dissipative damping of Rabi oscillations has been observed
experimentally \cite{Brune, Meek} and it has also been analyzed by
Bonifacio {\etal} who consider the evolution time as random variable
\cite{Boni}. In this letter we will show that a damping also
affects all the correlations functions which are involved in the
Bell inequality \cite{Asp} and in the separability criterion
\cite{Per, Wer} for the field and atomic internal variables. Can
this decoherence be interpreted as a ``which-way" information
effect? In this Letter we will try to throw light on these
questions.
\\Let us consider a two-level atom interacting with the field of
an optical cavity. In the rotating wave approximation (RWA) the
atom-field interaction is described by the Hamiltonian $\hbar
\epsilon \sin{k \hat{x}}(\hat{a}^{\dag} \hat{S}_{-}+\hat{a}$ $
 \hat{S}_{+})$ , where the usual spin $1/2$ operators refer to the
internal dynamics of the two-level atom, while $\hat{a}$ and
$\hat{a}^{\dag}$ are the annihilation and creation operators for
the photons of k-mode of the cavity standing wave and $\epsilon$
is the atom-field coupling constant. The atom enters the cavity
moving prevalently along a direction orthogonal to the  x-cavity
axis, and we assume that the atomic velocity  along this direction
is large enough to treat classically this component of the motion.
This interaction is at the heart of the optical Stern-Gerlach
effect \cite{sgo}, and it takes a form very similar to the
magnetic case introducing a new set of ``spin" operators
\begin{equation}\label{mu}
  (\hat{\mu}_{x},\,\hat{\mu}_{y},\,\hat{\mu}_{z})=(\frac{\hat{a}^{\dag} \hat{S}_{-}+\hat{a}
 \hat{S}_{+}}{2\sqrt{\hat{N}}},\,i\frac{\hat{a}^{\dag} \hat{S}_{-}-\hat{a}
 \hat{S}_{+}}{2\sqrt{\hat{N}}},\,\hat{S_{z}}),
\end{equation}
which satisfy in fact the algebra of a spin $1/2$,
$[\hat{\mu}_{x},\hat{\mu}_{y}]=i \hat{\mu}_{z}$ et cycl., where
$\hat{N}=\hat{a}^{\dag} \hat{a}+\hat{S}_{z}+\frac{1}{2}$. We
assume that the transverse spatial distribution of the incoming
atom is given by a packet of width $\Delta x_{0}$ narrow with
respect to the wavelength $\lambda$ of the resonant mode, and
centered near a nodal point of the cavity function. In these
conditions the sinusoidal mode function of the cavity
standing-wave can be approximated by the linear term \cite{Walls}.
In the resonance conditions the optical Stern-Gerlach Hamiltonian
reads
\begin{equation}\label{ham}
  \hat{H}=\frac{\hat{p}^{2}}{2 m}+\hbar\omega\hat{N}+\hbar
 \hat{\Omega}_{x}\hat{\mu}_{x},
\end{equation}
where $\hat{p}$ is the conjugate momentum of the position
observable $\hat{x}$, $m$ is the mass of the particle and
$\hat{\Omega}_{x}=2 \epsilon k \sqrt{\hat{N}} \hat{x}$. The
kinetic term of this Hamiltonian accounts for the atomic
translational degree of freedom along the x-direction. Consider
the one excitation initial configuration
\begin{equation}\label{psi0}
  \ket{\psi(0)}=\ket{e,0}\ket{\varphi(0)}=\frac{1}{\sqrt{2}}\left(\ket{\chi^{+}}+\ket{\chi^{-}}\right)
  \ket{\varphi(0)},
\end{equation}
where $\ket{\varphi(0)}$ is the initial ket of the translational
dynamics along the cavity axis, and we have expanded the ket
$\ket{e,0}\equiv \ket{e} \ket{0}$ in terms of the dressed states
$\ket{\chi^{\pm}}=\frac{1}{\sqrt{2}}(\ket{e,0}\pm\ket{g,1})$ which
are eigenstates of the excitation number operator,
$\hat{N}\ket{\chi^{\pm}}=\ket{\chi^{\pm}}$, and of the interaction
energy,
$\hat{\mu}_{x}\ket{\chi^{\pm}}=\pm\frac{1}{2}\ket{\chi^{\pm}}$. As
usually, $\ket{e}$ and $\ket{g}$ indicate the upper and the lower
states of the internal atomic dynamics, while $\ket{n}$ is a field
number state. Using these relations and the evolution operator
$\hat{U}(t,0)=e^{-i\hat{H}t/\hbar}$ we finally obtain the state of
the entire system at time $t\leq T$, (T is the cavity flight time)
\begin{equation}\label{psit}
  \ket{\psi(t)}=\frac{1}{\sqrt{2}}\left(\ket{\phi^{+}(t)}\ket{\chi^{+}}+
  \ket{\phi^{-}(t)}\ket{\chi^{-}}\right),
\end{equation}
where
\begin{equation}\label{phit}
  \ket{\phi^{\pm}(t)}=\exp(-i\frac{t}{2 m \hbar}\hat{p}^{2})\exp(\mp\frac{i}{2\hbar}a t^{2}\hat{p})
  \exp(\mp\frac{i}{\hbar}m\,a\,t\,\hat{x})\ket{\varphi(0)}
\end{equation}
account for the splitting of the translational state into two
parts, which go away in opposite sides, with a mean acceleration
$a=\hbar k \varepsilon/m$. In fact, in the p-representation we
have
\begin{equation}\label{phipt}
  \phi^{\pm}(p,t)=\braket{p\,}{\phi^{\pm}(t)}=\varphi(p\pm m\,a\,t,0)
  \exp\left[-i\frac{p\,t}{2\hbar}(\frac{p}{m}\mp \,a\, t)\right].
\end{equation}
Using the orthonormality of the dressed states
$\left|\chi^{\pm}\right>$ , the atomic momentum distribution is
\begin{equation}\label{psimod}
  \abs{\psi(p,t)}^{2}=\frac{1}{2}\left[\abs{\phi^{+}(p,t)}^{2}+\abs{\phi^{-}(p,t)}^{2}\right].
\end{equation}
\begin{figure}[t]
\includegraphics{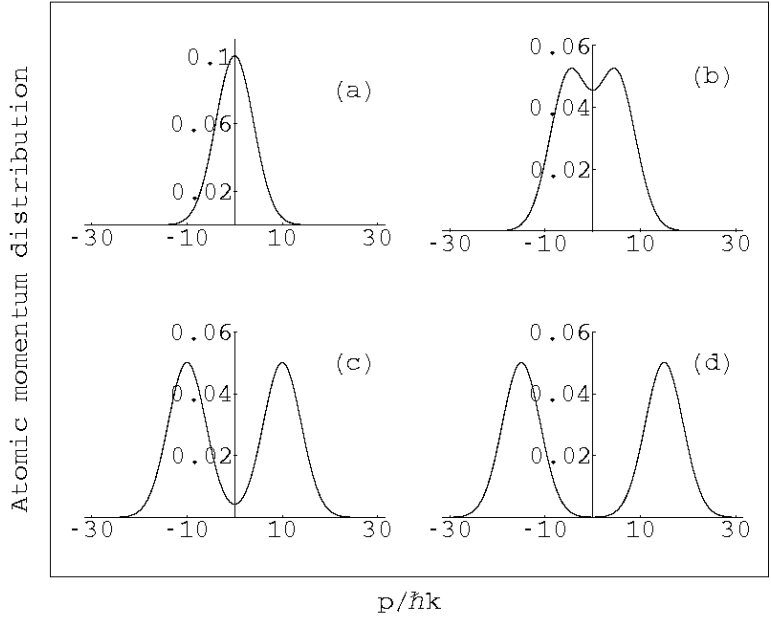} \caption{Momentum distribution
(Eq.(\ref{psimod})) for a two-level atom
 entering the cavity in the state $\ket{\varphi(0)}\ket{e}$
 and interacting with the vacuum state $\ket{0}$
 of the cavity field; $\varphi(p,0)$
 is a Gaussian packet of minimum uncertainty,
 with zero mean value of $\hat{p}$
 and $\Delta p_{0}/\hbar k = 25/2 \pi$ , corresponding to $\Delta x_{0}/\lambda=1/50$ .
 Curve (a) describes the initial distribution at time $t=0$.
 Curves (b), (c) and (d) refer to the interaction times $\epsilon T=5$, $\epsilon T=10$
 and $\epsilon T=15$, respectively, where $\epsilon=10^{8} sec^{-1}$. The values of the other
 parameters are $m=10^{-26}$ kg, $\lambda=10^{-5}$ meter.} \label{fig1}
\end{figure}
This probability density, shown in Fig.\ref{fig1} for some values
of the interaction time, displays the well known optical
Stern-Gerlach effect. The mechanical action of the light causes
the splitting of the atomic packet into two peaks which become
distinguishable for interaction time sufficiently large (a few
period of the Rabi frequency). This behaviour owns some
interesting and quite surprising features. The exchange of
momentum (through the photons exchange) between the atom and the
cavity walls cannot be a full random process since, in this case,
we would expect a single peak, centered in the initial mean value
of $\hat{p}$. On the contrary, both the Fig.\ref{fig1} and the
Eqs.(\ref{psit}) and (\ref{phit}) suggest that this exchange can
follow, in a single experiment, only one of two mutually exclusive
quantum paths. The momentum exchange behaves as only one cavity
side or the other should be involved with the same probability
(because of the particular initial state), and the Rabi
oscillations do emerge as a consequence of a coherent
superposition of these two quantum paths. According to this
scenario, it has been recently shown \cite{Cus} that the one-sided
atomic deflection accompanies with the so-called atomic coherent
trapping \cite{cohtrap}, in which the atomic internal population
is trapped into its initial value. In this extreme case the
particular initial configuration will cause the disappearing of
one of the two quantum paths, and the Rabi oscillations are absent
since the beginning. Certainly, the which-way information on these
two possibilities cannot be acquired by looking at the walls of
the cavity, which are absent from the model (\ref{ham}) (this is
equivalent to assume an infinite mass for the walls). However, we
can obtain an indirect which-path information by looking at the
momentum distribution of the deflected atom. For interaction times
sufficiently large, the two peaks of the atomic momentum are
distinguishable (see curves (c) and (d) of Fig.\ref{fig1}), and,
in the single experiment, one may be acquainted with the quantum
path the system has followed. At the same time, the interference
effects go to zero, as the damping of the Rabi oscillations shows
(see Fig.\ref{fig2}).\\ The correlations between the internal and
the translational atomic dynamics as displayed  in Eq.(\ref{psit})
lead to a drastic change in the time behavior of the internal
variables with respect to the usual Jaynes-Cummings model, in
which the atomic external variables are disregarded. Let $P(e,t)$
indicate the probability of finding the atom in the excited state
at time \emph{t},
\begin{equation}\label{pet}
  P(e,t)=\frac{1}{2}\left[1+\text{Re}\left(\braket{\phi^{-}(t)}{\phi^{+}(t)}\right)\right],
\end{equation}
where the relation
$\hat{\mu}_{z}\ket{\chi^{\pm}}=\frac{1}{2}\ket{\chi^{\mp}}$ has
been utilized. For example, if the initial translational state is
given by a Gaussian distribution of minimum uncertainty $\Delta
x_{0}\Delta p_{0}=\hbar/2$, centered in $x_{0}$  and with zero
mean velocity along the cavity axis, for the scalar product of
Eq.(\ref{pet}) we have
\begin{equation}\label{scal}
  \braket{\phi^{-}(t)}{\phi^{+}(t)}= e^{-i\,\Omega\,t}
 \exp\left\{-\frac{[x^{+}(t)-x^{-}(t)]^{2}}{8 \Delta x_{0}^{2}}-\frac{[p^{+}(t)-p^{-}(t)]^{2}}{8 \Delta
 p_{0}^{2}}\right\},
\end{equation}
where $x^{\pm}(t)=x_{0}\mp a \,t^{2}/2$, $p^{\pm}(t)=\mp
  m\,a\,t$ and $\Omega=2\,m \,a\,x_{0}/\hbar$.
\begin{figure}
\includegraphics{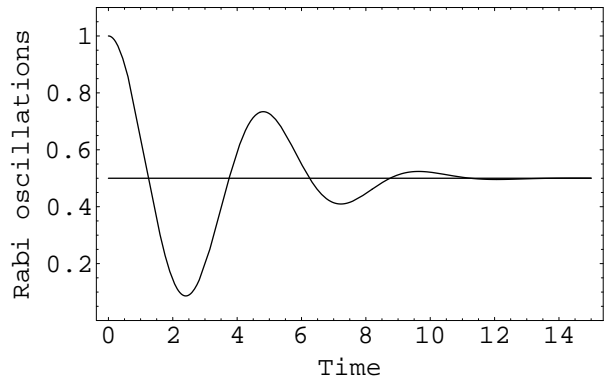} \caption{Time evolution of the atomic
population (Eq. (\ref{pet})).
 The time is in $\epsilon^{-1}$ units and $x_{0}=\frac{\lambda}{10}$. The values of the other parameters are as in Fig.\ref{fig1}.
 A similar damping affects the correlation functions of Eqs.(\ref{Mrho})  and (\ref{lamda}), which state the
 conditions for the violation of the Bell's inequality and for the separability,
 respectively.}
\label{fig2}
\end{figure}
As Eq.(\ref{scal}) shows, the scalar product between the two
translational components $\ket{\phi^{\pm}(t)}$ depends on the
distance in the phase space of the same components. When the
distance is sufficiently large the scalar product goes to zero,
the two paths become distinguishable and their interference
effects vanish. We wish to outline that this non dissipative
damping of the Rabi oscillations shown in Fig.\ref{fig2}, may be
considered as an intrinsic decoherence effect: It originates
inside the optical Stern-Gerlach model of Eq.(\ref{ham}), without
claiming any sort of reservoir action. It is to notice that the
modulus of the scalar product (\ref{scal}) essentially measures
the \emph{visibility V} (of the Rabi oscillations) as given by
Englert \cite{Engl}. We recall that, in our case, the
translational dynamics plays the role of the \emph{detector},
i.e., with the notation of Ref.\cite{Engl},
$\rho_{D}^{(i)}=\ket{\varphi(0)}\bra{\varphi(0)}$,
$\rho_{D}^{(+)}=\ket{\phi^{+}}\bra{\phi^{+}}$ and
$\rho_{D}^{(-)}=\ket{\phi^{-}}\bra{\phi^{-}}$. Similarly it is
possible to obtain the complementary quantity
\begin{equation}\label{disting}
  \emph{D}=\left(1-\abs{\braket{\phi^{+}}{\phi^{-}}}^{2}\right)^{\frac{1}{2}}.
\end{equation}
The \emph{distinguishability D} assumes a very simple form because
the initial state of the detector is a pure state. As pointed by
Englert, ``the ways cannot be distinguished at all if
$\emph{D}=0$" (when \emph{V}=1) ``and they can be held apart
completely if $\emph{D}=1$" (when \emph{V}=0). Due to the purity
of the initial state, the equality
\begin{equation}\label{disug}
  \emph{D}^{2}+\emph{V}^2
  =1-\abs{\braket{\phi^{+}}{\phi^{-}}}^{2}+\abs{\braket{\phi^{+}}{\phi^{-}}}^{2}=1
  \nonumber
\end{equation}
is satisfied at any time. In this contest the appearance of the
damping of the Rabi oscillations can be related to a
\emph{visibility} degradation. On the other hand, since the
quantum paths become distinguishable and a which-way information
is accessible, the quantum nature of the atom-field correlations
is vanishing. Consider the reduced density operator describing the
field and the atomic internal dynamics. Giving up the information
about the external variables by tracing the density operator of
the state Eq.(\ref{psit}) over the translational variables we get
\begin{equation}\label{rho}
  \rho=\frac{1}{2}\;\left[\,\ket{\chi^{+}}\bra{\chi^{+}}+\ket{\chi^{-}} \bra{\chi^{-}}+
  \left(\braket{\phi^{-}(t)}{\phi^{+}(t)}\ket{\chi^{+}}\bra{\chi^{-}}+h.c.\right)\,\right].
\end{equation}
We now investigate the nature of the atom-field correlations in
terms of the Bell's inequality. Because of its simplicity, we
consider the Horodecki family formulation \cite{Hor1}, which is
equivalent to the standard Clauser, Horne, Shimony, Holt, (CHSH)
formulation \cite{Cla}, when a bipartite system of spin $1/2$ is
involved, as in our case. The test reads: \emph{A density matrix
$\rho$ describing a system composed by two spin $1/2$  subsystems
violates some Bell's inequality in the CHSH formulation if and
only if the relation $M(\rho)>1$ is satisfied}. The quantity
$M(\rho)$  can be defined as follows. Consider the $3\times3$
matrix $T_{\rho}$ with coefficients
$t_{n,m}=tr(\rho\,\sigma_{n}\otimes \sigma_{m})$, where
$\sigma_{n}$ are the standard Pauli matrices. Diagonalizing the
symmetric matrix $U_{\rho}=T_{\rho}^{T}\cdot T_{\rho}$
($T_{\rho}^{T}$ is the transpose of $T_{\rho}$), and denoting two
greater eigenvalues of $U_{\rho}$ by $\lambda_{1}$ and
$\lambda_{2}$, then $M(\rho)=\lambda_{1}+\lambda_{2}$. In our case
we find
\begin{equation}\label{Mrho}
  M(\rho)=1+\left[\text{Im}\left(\braket{\phi^{-}(t)}
  {\phi^{+}(t)}\right)\right]^{2}.
\end{equation}
Looking at the Eqs. (\ref{psi0}) and (\ref{scal}) it is evident
that our system, which is initially non correlated, becomes, as
the atom-field interaction time increases, non-locally correlated.
Quite interestingly, at the same time the atomic external
dynamics, which makes accessible the which-way information, starts
to damp this non-locality. Since the violation of the Bell's
inequality is just a sufficient condition for non-locality, it may
be useful to consider the separability \cite{Per, Wer} of the
density matrix because it is a sufficient condition for both the
locality and the classical correlation. After a few periods of
Rabi oscillations the damping factor involved in the scalar
product (\ref{scal}), determines the separability of the system in
the sense that it is possible to set it in the form
$\rho=\sum_{r=1}^{\infty}p_{r}\,W_{r}^{(1)}\otimes W_{r}^{(2)}$,
where $W_{r}^{(1)}$  and $W_{r}^{(2)}$  are states corresponding
to the single subsystems and $p_{r}$ are probabilities
($\sum_{r=1}^{\infty}p_{r}=1$). To test this possibility in a more
quantitative way we consider the eigenvalues of the partial
transpose \cite{Per} $\sigma_{\rho}$  of the matrix $\rho$. A
necessary and sufficient condition for separability reads
\cite{Per, Hor2}: \emph{$\rho$ is separable if and only if all the
eigenvalues of $\sigma_{\rho}$ are non-negative}. In our case the
test writes
\begin{equation}\label{lamda}
  \abs{\,\text{Im}\left(\braket{\phi^{-}(t)}
  {\phi^{+}(t)}\right)}\leq 0.
\end{equation}
Because of the correlations with the atomic external variables the
classicality of the correlations between the field and the atomic
internal variables is recovered, as explicitly shown by the
diagonal form
$\rho=\frac{1}{2}(\ket{e}\bra{e}\otimes\ket{0}\bra{0}
+\ket{g}\bra{g}\otimes\ket{1}\bra{1})$ that the density matrix
(\ref{rho}) assumes after a few Rabi periods (when
$\braket{\phi^{-}(t)}{\phi^{+}(t)} \rightarrow 0$). It is
intriguing the fact that the which-way information can be
accessible just when the quantum nature of the atomic
translational dynamics  is taken into account.\\
\\
In conclusion, the interaction between the cavity field and the
internal atomic variables can pursue two mutually exclusive
quantum paths. Due to the correlation with the atomic external
variables, these two paths cause the well known optical
Stern-Gerlach effect, that is, the atomic translational packet
splits into two parts. When these are sufficiently separate, a
which way information becomes accessible, and all the correlations
functions involved in the Bell's inequality, in the separability
criterion, and in the Rabi oscillations go exponentially to zero,
showing a propensity towards a classical behavior. In a
forthcoming paper we will use the same model to prove how the
atomic translational effects make more difficult the violation of
Bell's inequality for massive particles.
%


\end{document}